\documentclass[preprint,preprintnumbers,amsmath,amssymb]{revtex4}%

\usepackage{graphicx}
\usepackage{dcolumn}
\usepackage{bm}
\usepackage{amsmath}
\usepackage{epstopdf}
\usepackage{amsfonts}
\usepackage{amssymb}%
\providecommand{\U}[1]{\protect\rule{.1in}{.1in}}
\newcommand{\be}{\begin{equation}}
\newcommand{\en}{\end{equation}}
\newcommand{\bea}{\begin{eqnarray}}
\newcommand{\ena}{\end{eqnarray}}

\begin{document}

\title{Quasinormal modes of the BTZ black hole under scalar perturbations with a non-minimal coupling: Exact spectrum}

\author{Grigoris Panotopoulos}

\email{grigorios.panotopoulos@tecnico.ulisboa.pt}

\affiliation{Centro de Astrof\'{\i}sica e Gravita{\c c}{\~a}o, Instituto Superior T\'ecnico-IST,Universidade de Lisboa-UL, Av. Rovisco Pais 1, 1049-001 Lisboa, Portugal}

\date{\today}

\begin{abstract}
We perturb the non-rotating BTZ black hole with a non-minimally coupled massless scalar field, and we compute the quasinormal spectrum exactly.
We solve the radial equation in terms of hypergeometric functions, and we obtain an analytical expression for the quasinormal frequencies. In addition,
we compare our analytical results with the 6th order semi-analytical WKB method, and we find an excellent agreement. The impact of the nonminimal coupling as well as of the cosmological constant on the quasinormal spectrum is briefly discussed.
\end{abstract}

\maketitle

\section{Introduction}

Black holes (BHs), a generic prediction of Einstein's General Relativity (GR), are way more than just mathematical objects. After Hawking's seminal work \cite{hawking1,hawking2} in which it was shown that BHs emit radiation from their horizon, these objects have attracted a
lot of attention over the last decades, as they comprise an excellent laboratory to study and understand several aspects of gravitational theories.
In fact it is often said that BHs are the simplest macroscopic bodies in Nature,
since they are uniquely characterized by their few parameters only, such as the mass, the charge, and the rotation speed. This statement is due to the theoretical paradigm of the no-hair conjecture \cite{misner}, although counter examples do exist \cite{counter1,counter2}. 
 
When BHs are perturbed the geometry of spacetime undergoes dumbed oscillations. The work of \cite{wheeler} marked the birth of BH perturbations, and it was later extended by \cite{zerilli1,zerilli2,zerilli3,moncrief,teukolsky}. The state-of-the art in BH perturbations is summarized in Chandrasekhar's monograph \cite{monograph}.
Quasinormal modes (QNM) with a non-vanishing imaginary part, since they do not depend on the initial conditions, carry unique information about the few BH parameters. After the LIGO historical direct detections of gravitational waves \cite{ligo1,ligo2,ligo3}, that offer us the strongest evidence so far that BH exist and merge, QNM of black holes are now more relevant than ever. By observing the BH quasinormal spectrum, that is frequencies and damping rates, eventually we will be able to prove or falsify the theoretical paradigm of the no-hair conjecture. 
QNM of BHs have been extensively studied in the literature. For a review on the subject see \cite{review1}, and for a more recent one \cite{review2}.

Gravity in 1+2 dimensions is special, and it has attracted a lot of interest for several reasons. First, due to the deep connection to a Yang-Mills theory with only the Chern-Simons term \cite{CS,Witten:1988hc,Witten:2007kt}. Furthermore, three-dimensional BHs have thermodynamic properties closely analogous to those of realistic (1+3)-dimensional black holes: they have horizons, entropy, and they radiate at a Hawking temperature. Due to the absence of propagating degrees of freedom they offer us the possibility to understand four-dimensional BHs in a simpler framework. 
Of particular interest is the Ba{\~n}ados, Teitelboim and
Zanelli (BTZ) black-hole \cite{BTZ,Banados:1992gq}, which lives in three dimensions, and the presence of a negative cosmological constant
is crucial for the existence of the black hole. A complete review on BTZ black hole can be found in \cite{review}. 

Computing the QNM frequencies in an analytical way is possible in a few cases only \cite{cardoso2,exact,potential,ferrari1,zanelli,fernando1,fernando2}, while in most of the cases either some numerical scheme \cite{numerics1,numerics2,numerics3} or semi-analytical methods are employed, such as the well-known from standard quantum mechanics WKB method used extensively in the literature \cite{wkb1,wkb2,paper1,paper2,paper3,paper4,paper5,paper6,paper7,paper8}.
Clearly, scalar perturbations is the simplest case to consider. What is more, the scalar field that perturbs the BH is usually taken to be a canonical massless one. However, other possibilities have been considered too, such as charged scalar field in the case of charged BHs \cite{fernando2}, a massive scalar field \cite{exact,zanelli}, or a non-minimally coupled scalar field \cite{zanelli, mann}.

The QN modes of a non-rotating BTZ BH for a massless canonical scalar field were computed in \cite{cardoso2}, while the QN frequencies of a rotating BTZ BH for a massive scalar field were computed in \cite{exact}. Furthermore, the QN spectrum for massless topological BHs in four and higher dimensions were computed in \cite{zanelli}, where a non-minimal coupling was introduced as well, while in \cite{mann} the falloff behaviour of conformally invariant scalar waves in asymptotically anti-de Sitter backgrounds was investigated. In the present work we perturb the non-rotating BTZ black hole
\cite{BTZ,Banados:1992gq} with a non-minimally coupled massless scalar field, and we solve the radial equation exactly to obtain an analytical expression for the quasinormal spectrum in the strong coupling regime, $\xi > 1/6$ (see the discussion below), adding thus in the literature one more exact analytical calculation. Therefore it is the goal of this article to study the propagation of a probe scalar field in the following gravitational background
\begin{equation}
ds^2 = -\left(-M + \frac{r^2}{l^2} \right) dt^2 + \frac{1}{\left(-M + \frac{r^2}{l^2} \right)} dr^2 + r^2 d \phi^2
\end{equation}
with $M$ being the mass of the black hole, and $l$ being the negative cosmological constant, $\Lambda = -1/l^2$,
and investigate the stability of the BTZ BH under scalar perturbations with a non-vanishing non-minimal coupling to gravity.  

Our work is organized as follows: After this introduction, the nonminimally coupled scalar field and its wave equation are discussed in section 2, in which we also present the effective potential. In the third section we solve the radial equation in terms of hypergeometric functions, and we
provide an analytical expression for the quasinormal spectrum in section 4. Finally, we conclude our work in the last section.

\section{Scalar perturbations}

Next we consider in the above gravitational background a probe massless scalar field with a nonzero coupling $\xi$ to the Ricci scalar described by the action
\begin{equation}
S = \frac{1}{2} \int d^3x \sqrt{-g} \Bigl[  \partial^{\mu}\Phi \partial_{\mu}\Phi + \xi R_3 \Phi^2 \Bigl]
\end{equation}
In the given BTZ spacetime the wave equation of the scalar field reads \cite{coupling,kanti2,Pappas:2016ovo}
\begin{equation}
\frac{1}{\sqrt{-g}} \partial_\mu (\sqrt{-g} g^{\mu \nu} \partial_\nu) \Phi = \xi R_3 \Phi
\end{equation}
where the nonminimal coupling is taken to be positive, and $R_3=-6/l^2$ is the constant
Ricci scalar of the BTZ background. Making the ansatz
\begin{equation}\label{separable}
\Phi(t,r,\phi) = e^{-i \omega t} R(r) e^{i m \phi}
\end{equation}
we obtain an ordinary differential equation for the radial part
\begin{equation}
R'' + \left( \frac{1}{r} + \frac{f'}{f} \right) R' + \left( \frac{\omega^2}{f^2} - \frac{m^2}{r^2 f} + \frac{6 \xi}{l^2 f} \right) R = 0
\end{equation}
where $f(r)=-M+r^2/l^2$ is the metric function in the BTZ BH solution. Note that the nonzero coupling to the scalar curvature can be interpreted as a mass term when the cosmological constant is positive. Here, however, since the cosmological constant is negative, the mass term enters with the wrong sign.
To see the effective potential that the scalar field feels we define new variables as follows
\begin{eqnarray}
R & = & \frac{\psi}{\sqrt{r}} \\
x & = & \int \frac{dr}{f(r)}
\end{eqnarray}
where we are using the so-called tortoise coordinate $x$ given by
\begin{equation}
x = \frac{l^2}{2 r_H} \ln \left( \frac{r-r_H}{r+r_H} \right)
\end{equation}
with $r_H=l \sqrt{M}$ being the event horizon,
and recast the equation for the radial part into a Schr{\"o}dinger-like equation of the form
\begin{equation}
\frac{d^2 \psi}{dx^2} + (\omega^2 - V(x)) \psi = 0
\end{equation}
Therefore we obtain for the effective potential the expression
\begin{equation}
V(r) = f(r) \: \left(-\frac{6 \xi}{l^2} + \frac{m^2}{r^2}+\frac{f'(r)}{2 r}-\frac{f(r)}{4 r^2} \right)
\end{equation}
and as a function of the radial coordinate can be seen in Figures 1 and 2
for three different values of the coupling $\xi=0.25, 0.5, 0.75$ and three different values of the cosmological constant $l=4, 5, 6$, respectively.
The maximum of the potential is located at 
\begin{equation}
r_0 = \frac{l \sqrt{M}}{(24 \xi - 3)^{1/4}}
\end{equation}
while the maximum value of the potential $V_{max} = V(r_0)$ is computed to be
\begin{equation}
V_{max} = \frac{M}{2 l^2} \; \frac{3-\sqrt{24 \xi - 3}+12 \xi (\sqrt{24 \xi - 3}-2)}{\sqrt{24 \xi - 3}}
\end{equation}
It is very easy to verify that the maximum of the potential increases with $M$ and $\xi$ and decreases with $l$. Therefore when the cosmological constant increases destabilizes the system, while the opposite holds for the mass of the black hole.

\begin{figure}
\centering
\includegraphics[scale=1.2]{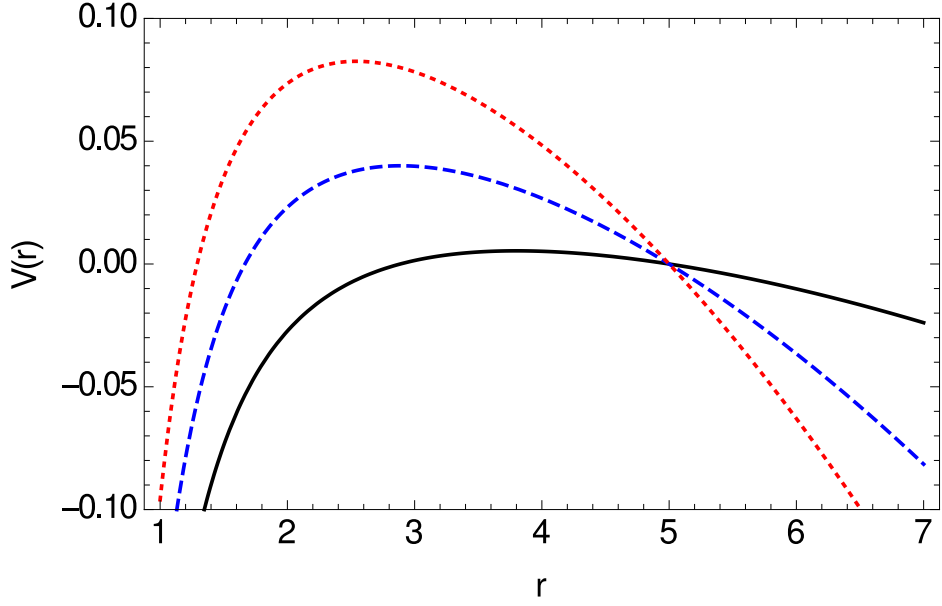}
\caption{Effective potential for $m=0, l=5, M=1$ and for $\xi=0.25$ (solid black curve), $\xi=0.5$ (dashed blue curve) and $\xi=0.75$ (dotted red curve).}
\end{figure}

\begin{figure}
\centering
\includegraphics[scale=1.2]{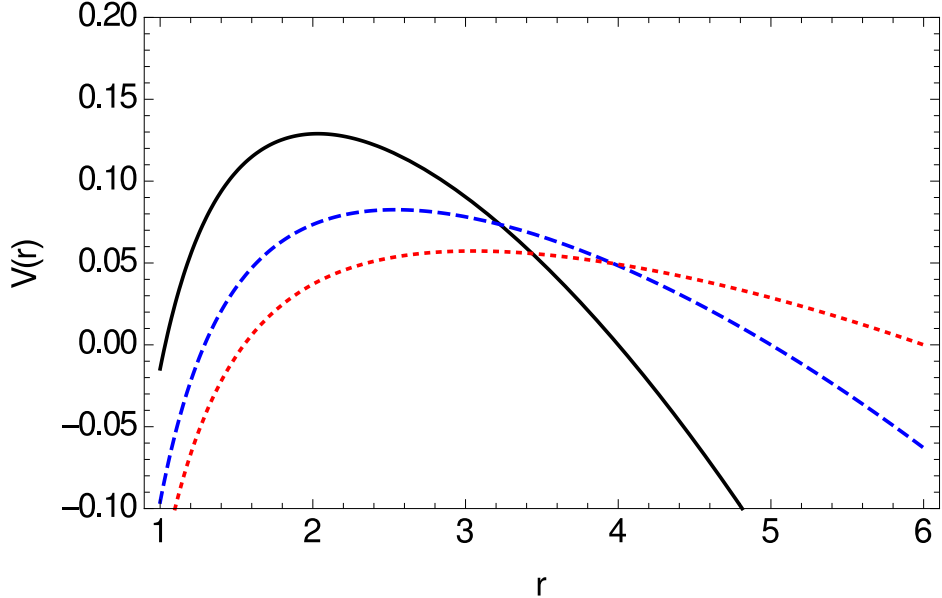}
\caption{Effective potential for $m=0, \xi=0.75, M=1$ and for $l=4$ (solid black curve), $l=5$ (dashed blue curve) and $l=6$ (dotted red curve).}
\end{figure}

To complete the formulation of the physical problem, we must also impose the appropriate boundary conditions at horizon ($r \simeq r_H$) and at infinity ($r \rightarrow \infty$), which are the following \cite{superradiance, ferrari2}
\begin{equation}
\psi(x) \rightarrow
\left\{
\begin{array}{lcl}
A e^{-i \omega x} & \mbox{ at horizon } & (x \rightarrow - \infty) \\
&
&
\\
 C_- e^{i k x} & \mbox{ at infinity } & (x \rightarrow 0^-)
\end{array}
\right.
\end{equation}
where $A$ is an arbitrary constant, while $k, C_-$ are constants that can be computed in terms of the parameters of the model at hand. The purely in-going wave physically means that nothing can escape from the horizon, while the purely out-going wave corresponds to the requirement that no radiation is incoming from infinity \cite{ferrari2}. The latter requirement allows us to obtain an infinite set of discrete complex numbers, $\omega=\omega_R + \omega_I i$, which are precisely the QN frequencies of the black hole. Given the time dependence of the probe scalar field $\Phi \sim e^{-i \omega t}$, it is clear that unstable modes correspond to $\omega_I > 0$, while stable modes correspond to $\omega_I < 0$. In the second case the real part of the mode $\omega_R$ determines the period of the oscillation, $T=2 \pi/\omega_R$, while the imaginary part $|\omega_I|$ describes the decay of the fluctuation at a time scale $t_D=1/|\omega_I|$.

\section{Solution of the full radial equation in terms of hypergeometric functions}

To find the solution of the full radial equation we introduce the dimensionless parameter $z=1-r_H^2/r^2$ which takes values
between 0 (horizon, $r \rightarrow r_H$) and 1 (far-field region, $r \gg r_H$). The new differential equation with respect to $z$ takes the form
\begin{equation}
z (1-z) R_{zz} + (1-z) R_z \ +
\left( \frac{A}{z} + \frac{B}{-1+z} - C \right) R = 0
\end{equation}
where the three constant are given by
\begin{eqnarray}
A & = & \frac{l^4 \omega^2}{4 r_H^2} \\
B & = & -\frac{3 \xi}{2} \\
C & = & \frac{l^2 m^2}{4 r_H^2}
\end{eqnarray}
The last differential equation can be recast in the form of the Gauss' hypergeometric equation by
removing the poles in the last term making the ansatz
\begin{equation}
R = z^\alpha (1-z)^\beta F
\end{equation}
where now $F$ satisfies the following differential equation
\begin{equation}
z (1-z) F_{zz} + [1+2 \alpha - (1+2 \alpha+2 \beta) z] F_z +
\left( \frac{\bar{A}}{z} + \frac{\bar{B}}{-1+z} - \bar{C} \right) F = 0
\end{equation}
and the new constants are given by
\begin{eqnarray}
\bar{A} & = & A + \alpha^2 \\
\bar{B} & = & B + \beta - \beta^2 \\
\bar{C} & = & C+(\alpha+\beta)^2
\end{eqnarray}

To remove the poles at $z=0$ and $z=1$ we demand that $\bar{A}=0=\bar{B}$. The parameter $\alpha$ is computed to be
\begin{equation}
\alpha = - i \frac{l^2 \omega}{2 r_H}
\end{equation}
while for $\beta$ there are two cases. First, in the low coupling regime, $6 \xi < 1$, the determinant of the algebraic equation of second degree for $\beta$ is positive, and $\beta$ is real given by
\begin{equation}
\beta  =  \frac{1 - \sqrt{1 - 6 \xi}}{2}
\end{equation}
In this case the quasinormal spectrum is given by eq. (18) of \cite{exact} by the replacement $\mu \rightarrow -6 \xi$. If, however, $6 \xi > 1$ (let us call it the strong coupling regime), $\beta$ is complex and it is given by
\begin{equation}
\beta  =  \frac{1 + i\sqrt{6 \xi - 1}}{2}
\end{equation}
This is the case we shall consider in the rest of the discussion.

Finally we obtain the hypergeometric equation
\begin{equation}
z (1-z) F_{zz} + [c-(1+a+b) z] F_z - ab F = 0
\end{equation}
with parameters $a,b,c$ given by
\begin{eqnarray}
c & = & 1+2 \alpha \\
a & = &  \alpha + \beta + i \sqrt{C} \\
b & = & \alpha + \beta - i \sqrt{C}
\end{eqnarray}
Note that the parameters $a,b,c$ satisfy the condition $c-a-b=1-2 \beta$.
Therefore the general solution for the radial part is given by \cite{handbook}
\begin{equation}
R(z) = z^\alpha (1-z)^\beta \Bigl[ C_1 F(a,b;c;z) + C_2 z^{1-c} F(a-c+1,b-c+1;2-c;z) \Bigl]
\end{equation}
where $C_1,C_2$ are two arbitrary coefficients,
and the hypergeometric function can be expanded in a Taylor series as follows \cite{handbook}
\begin{equation}
F(a,b;c;z) = 1 + \frac{a b}{c} \: z+ \cdots
\end{equation}
Setting $C_2=0$ and for the choice for $\alpha=-i (l^2 \omega)/(2 r_H)$ we recover the purely in-going solution close to the horizon,
$R \sim (r-r_H)^\alpha$.
Therefore in the following we consider the first solution only, namely
\begin{equation}
R(z) = D z^\alpha (1-z)^\beta F(a,b;c;z)
\end{equation}
where now we have replaced $C_1$ by $D$.

\section{Exact spectrum}

In order to reveal the behaviour of radial part in the far-field region
(where $z \rightarrow 1$) we use the transformation \cite{handbook}
\begin{equation}
\begin{split}
F(a,b;c;z) = \ &\frac{\Gamma(c) \Gamma(c-a-b)}{\Gamma(c-a) \Gamma(c-b)} \times \:
\\
&F(a,b;a+b-c+1;1-z) \ +
 \\
(1-z)^{c-a-b} &\frac{\Gamma(c) \Gamma(a+b-c)}{\Gamma(a) \Gamma(b)} \times \:
\\
&F(c-a,c-b;c-a-b+1;1-z)
\end{split}
\end{equation}
and therefore the radial part as $z \rightarrow 1$ reads
\begin{equation}
\begin{split}
R_{FF} \simeq  \frac{D (1-z)^\beta \Gamma(1+2 \alpha) \Gamma(1-2 \beta)}{\Gamma(1+\alpha-\beta-i \sqrt{C}) \Gamma(1+\alpha-\beta+i \sqrt{C})}
\\
+ \frac{ D (1-z)^{1-\beta} \Gamma(1+2 \alpha) \Gamma(-1+2 \beta)}{\Gamma(\alpha+\beta-i \sqrt{C}) \Gamma(\alpha+\beta+i \sqrt{C})}
\end{split}
\end{equation}
Upon defining two new constants
\begin{eqnarray}
D_- & = & D \: \frac{\Gamma(1+2 \alpha) \Gamma(1-2 \beta)}{\Gamma(1+\alpha-\beta-i \sqrt{C}) \Gamma(1+\alpha-\beta+i \sqrt{C})} \\
D_+ & = & D \: \frac{\Gamma(1+2 \alpha) \Gamma(-1+2 \beta)}{\Gamma(\alpha+\beta-i \sqrt{C}) \Gamma(\alpha+\beta+i \sqrt{C})}
\end{eqnarray}
and since $z=1-(r_H/r)^2$, the radial part $R(r)$ for $r \gg r_H$ can be written down as follows
\begin{equation}
R_{FF} \simeq D_- \left( \frac{r}{r_H} \right)^{-2 \beta} + D_+ \left( \frac{r}{r_H} \right)^{2 \beta-2}
\end{equation}
where the $D_-$ term corresponds to the in-going wave, while the $D_+$ term corresponds to the out-going one \cite{Fernando:2004ay,PR}. 

The discussion of section 2 on the boundary conditions at infinity is valid for asymptotically flat spacetimes. In the present work, however, since there is a negative cosmological constant the correct approach is to require that the whole solution vanishes at infinity, as it has been already done in previous works \cite{ex1,ex2}. Therefore, in the last step of the calculation we require that $D_-=0$ and $D_+=0$, which happens when the Gamma functions in the denominators have a pole. The first condition implies
\begin{equation}
-n = 1 + \alpha - \beta \pm i \sqrt{C}
\end{equation}
while the second condition implies
\begin{equation}
-n = \alpha + \beta \pm i \sqrt{C}
\end{equation}
where $n=0,1,2,...$ is the overtone number.
Using the expressions for $C, \alpha, \beta$ obtained in the previous section, it is easy to verify that both conditions are met only when the quasinormal frequencies are given by the following expression
\begin{equation}
 \boxed{\omega_n = \left| \frac{|m|}{l} - \frac{\sqrt{M}}{l} \: \sqrt{6 \xi - 1} \right| - \frac{2 \sqrt{M}}{l} \left( n+\frac{1}{2} \right) i}
\end{equation}
which is the main result of this work. We wish to remark here that this results holds in the strong regime only, $\xi > 1/6$, and therefore the limit $\xi=0$ cannot be taken to reproduce the formula computed in \cite{cardoso2}. 

We see that the quasinormal modes have both real and imaginary part, with the latter being always negative, and therefore the modes are stable. Contrary to the results of previous works where it was found that both the mass of the scalar field and the non-minimal coupling enter into the imaginary part of the frequencies \cite{exact, zanelli}, we find here that $\xi$ in the strong regime enters into the real part. Therefore the non-minimal coupling does not affect the stability of the BH. In addition, compared to the canonical scalar field case \cite{cardoso2}, the presence of the non-minimal coupling modifies both the real and the imaginary part of the spectrum. Note the difference between the $(n+1)$ factor of \cite{cardoso2} in the imaginary part, and the $n+(1/2)$ factor obtained here. The presence of the cosmological constant and the mass of the BH remain the same as in \cite{cardoso2, exact}, $\omega_I \sim \sqrt{M}/l$ (see the formulas (41) and (42) below). As we already mentioned when we presented the effective potential as a function of the radial coordinate, $M$ stabilizes the BH while the cosmological constant destabilizes it. Finally, the angular momentum $m$ only affects the real part of the frequencies, while the overtone number only affects the imaginary part, precisely as in the $\xi=0$ case as well as in the $0 \leq \xi < 1/6$ case.

To confirm our analytical expression we have computed the QN frequencies using the popular semi-analytical WKB method \cite{wkb1,wkb2}. The QN modes
within the WKB approximation are given by
\begin{equation}
\omega^2 = V_0+(-2V_0'')^{1/2} \Lambda(n) - i \nu (-2V_0'')^{1/2} [1+\Omega(n)]
\end{equation}
where $n=0,1,2...$ is the overtone number, $\nu=n+1/2$, $V_0$ is the maximum of the effective potential, $V_0''$ is the second derivative of the effective potential evaluated at the maximum, while $\Lambda(n), \Omega(n)$ are complicated expressions of $\nu$ and higher derivatives of the potential evaluated at the maximum, and can be seen e.g. in \cite{paper2,paper7}. Here we have used the Wolfram Mathematica \cite{wolfram} code with WKB at any order from one to six presented in \cite{code}.

Our results, summarized in Table \ref{Firstset} below, show an excellent agreement between the numerical results and our explicit formula. For comparison, in Table \ref{Secondset} we show the quasinormal frequencies corresponding to the $\xi=0$ case \cite{cardoso2}
\begin{equation}
\omega_{\xi=0} = \frac{|m|}{l} - \frac{2 \sqrt{M}}{l} (n+1) i
\end{equation}
as well as to the $\xi=0.1$ (weak coupling regime, $0 \leq \xi < 1/6$) \cite{exact}
\begin{equation}
\omega_{\xi < 1/6} = \frac{|m|}{l} - \frac{2 \sqrt{M}}{l} \left(n+\frac{1}{2}+\frac{\sqrt{1-6 \xi}}{2} \right) i
\end{equation}
which reproduces the previous result for $\xi=0$. 

Finally, in Table \ref{Thirdset} we show the quasinormal modes for fixed BH mass, $M=1$, and nonminimal coupling, $\xi=0.5$, and three different values of the cosmological constant $l=1, 2, 3$. As we have already mentioned, when the cosmological constant increases the black hole becomes less stable, as it was expected since the imaginary part of the frequencies depends inversely proportional on $l$.

\begin{table}
\centering
\begin{tabular}{||c | c | c | c ||} 
\hline
  \multicolumn{4}{||c||}{$l=1, \, M=1, \, \xi=0.25$} \\
   \hline
    n & m=1 & m=2 & m=3 \\ [0.5ex] 
    \hline
   0  & 0.292893-1. i  & 1.29289-1. i & 2.29289 -1. i \\ 
    & (0.291073-0.999683 i) & (1.29148-0.999136 i) & (2.29191-0.999315 i) \\
   \hline
   1 & 0.292893-3. i & 1.29289-3. i & 2.29289-3. i \\
    & (0.290992-2.99989 i) & (1.29065-2.99935 i) & (2.29103-2.9991 i) \\  
   \hline
   2 &  & 1.29289-5. i & 2.29289-5. i \\  
    &  & (1.29048-4.99957 i) & (2.29067-4.99929 i)  \\
    \hline
   3 &  &  & 2.29289-7. i \\  
     &  &  & (2.29053-6.99945 i) \\
   \hline
   \hline
\multicolumn{4}{||c||}{$l=1, \, M=1, \, \xi=0.5$} \\
 \hline
     n & m=2 & m=3 & m=4 \\ [0.5ex] 
     \hline
  0  & 0.585786-1. i  & 1.58579-1. i & 2.58579-1. i  \\ 
  & (0.5857-0.999958 i) & (1.58574-0.999946 i) & (2.58576-0.999958 i) \\
  \hline
 1 & 0.585786-3. i & 1.58579-3. i & 2.58579-3. i  \\ 
 & (0.585679-2.99998 i) & (1.58568-2.99995 i) & (2.5857-2.99994 i) \\
 \hline
 2 & 0.585786-5. i & 1.58579-5. i & 2.58579-5. i \\ 
  & (0.585677-4.99999 i) & (1.58566-4.99997 i) & (2.58568-4.99995 i) \\
   \hline
   3 &  & 1.58579-7. i & 2.58579-7. i \\ 
    &  & (1.58566-6.99997 i) & (2.58566-6.99996 i) \\
  \hline
  4 &  &  & 2.58579-9. i \\ 
   &  &  & (2.58566-8.99997 i) \\
  \hline
  \multicolumn{4}{||c||}{$l=1, \, M=1, \, \xi=0.75$} \\
 \hline
     n & m=2 & m=3 & m=4 \\ [0.5ex] 
     \hline
 0  & 0.129171-1. i  & 1.12917-1. i & 2.12917-1. i \\ 
  & (0.129161-0.999998 i) & (1.12916-0.999985 i) & (2.12916-0.999987 i) \\
 \hline
 1 & 0.129171-3. i & 1.12917-3. i & 2.12917-3. i \\
  & (0.129161-3. i) & (1.12914-2.99999 i) & (2.12915-2.99998 i) \\  
 \hline
 2 & 0.129171-5. i & 1.12917-5. i & 2.12917-5. i \\  
  & (0.129162-4.99997 i) & (1.12914-4.99999 i) & (2.12914-4.99999 i) \\
 \hline
 3 &  & 1.12917-7. i & 2.12917-7. i \\ 
   &  & (1.12914-7. i) & (2.12914-6.99999 i) \\
 \hline
 4 &  &  & 2.12917-9. i \\ 
   &  &  & (2.12914-8.99999 i) \\
 \hline
\end{tabular}

\caption{Scalar QNMs of non-rotating BTZ black hole for $l=1=M$ and three different values of the non-minimal coupling $\xi=0.25, 0.5, 0.75$. 
$m,\,n$ are the angular momentum and overtone number, respectively. The values without the parenthesis are the exact QNMs, while the ones in the parenthesis are the values obtained using the 6th order WKB method.}
\label{Firstset}
\end{table}

\begin{table}
\centering
\begin{tabular}{||c | c | c | c ||} 
\hline
  \multicolumn{4}{||c||}{$l=1, \, M=1, \, \xi=0$} \\
   \hline
n & m=1 & m=2 & m=3 \\ [0.5ex] 
    \hline
   0  & 1-2 i  & 2-2 i & 3-2 i  \\
    \hline
   1 & 1-4 i & 2-4 i & 3-4 i \\
   \hline
   2 &  & 2-6 i & 3-6 i \\  
    \hline
   3 &  &  & 3-8 i \\  
   \hline
   \hline
\multicolumn{4}{||c||}{$l=1, \, M=1, \, \xi=0.1$} \\
 \hline
 n & m=1 & m=2 & m=3 \\ [0.5ex] 
    \hline
   0  & 1-1.63246 i  & 2-1.63246 i & 3-1.63246 i \\ 
   \hline
   1 & 1-3.63246 i  & 2-3.63246 i & 3-3.63246 i \\
   \hline
   2 &  & 2-5.63246 i & 3-5.63246 i \\  
    \hline
   3 &  &  & 3-7.63246 i \\  
   \hline
\end{tabular}

\caption{Scalar QNMs of non-rotating BTZ black hole for $l=1=M$ and two different values of the nonminimal coupling, namely $\xi=0$ (no coupling, upper part) and $\xi=0.1$ (weal coupling regime, $0 \leq \xi < 1/6$, lower part). 
$m,\,n$ are the angular momentum and overtone number, respectively.}
\label{Secondset}
\end{table}

\begin{table}
\centering
\begin{tabular}{||c | c | c | c ||} 
\hline
  \multicolumn{4}{||c||}{$l=1, \, M=1, \, \xi=0.5$} \\
   \hline
     n & m=2 & m=3 & m=4 \\ [0.5ex] 
     \hline
  0  & 0.585786-1. i  & 1.58579-1. i & 2.58579-1. i  \\ 
  \hline
 1 & 0.585786-3. i & 1.58579-3. i & 2.58579-3. i  \\ 
 \hline
 2 & 0.585786-5. i & 1.58579-5. i & 2.58579-5. i \\ 
  \hline
   3 &  & 1.58579-7. i & 2.58579-7. i \\ 
  \hline
  4 &  &  & 2.58579-9. i \\ 
  \hline
  \hline
  \multicolumn{4}{||c||}{$l=2, \, M=1, \, \xi=0.5$} \\
  \hline   
     n & m=2 & m=3 & m=4 \\ [0.5ex] 
     \hline
  0  & 0.292893-0.5 i  & 0.792893-0.5 i & 1.29289-0.5 i \\ 
  \hline
 1 & 0.292893-1.5 i & 0.792893-1.5 i & 1.29289-1.5 i  \\ 
 \hline
 2 & 0.292893-2.5 i & 0.792893-2.5 i & 1.29289-2.5 i \\ 
  \hline
   3 &  & 0.792893-3.5 i & 1.29289-3.5 i \\ 
    \hline
  4 &  &  & 1.29289-4.5 i \\ 
   \hline
   \hline
  \multicolumn{4}{||c||}{$l=3, \, M=1, \, \xi=0.5$} \\
 \hline  
n & m=2 & m=3 & m=4 \\ [0.5ex] 
     \hline
  0  & 0.195262-0.333333 i  & 0.528595-0.333333 i & 0.861929-0.333333 i \\ 
  \hline
 1 & 0.195262-1.0 i  & 0.528595-1.0 i & 0.861929-1.0 i  \\ 
 \hline
 2 & 0.195262-1.66667 i & 0.528595-1.66667 i & 0.861929-1.66667 i \\ 
  \hline
   3 &  & 0.528595-2.33333 i & 0.861929-2.33333 i \\ 
  \hline
  4 &  &  & 0.861929-3.0 i \\ 
  \hline
\end{tabular}

\caption{Scalar QNMs of non-rotating BTZ black hole for $M=1$, $\xi=0.5$ and three different values of the cosmological constant $l=1,2,3$. 
$m,\,n$ are the angular momentum and overtone number, respectively.}
\label{Thirdset}
\end{table}

\section{Conclusions}

To summarize, in this article we have studied the quasinormal spectrum of a non-rotating BTZ black hole by analysing the propagation of a probe massless non-minimally coupled scalar field, which perturbs the BTZ black hole background. Solving the radial equation in terms of the hypergeometric functions we have obtained an exact analytical expression for the quasinormal frequencies. Compared to the case of a canonical scalar field, we find that the presence of the non-minimal coupling modifies both the real and the imaginary part of the frequencies, although it does not enter into the latter explicitly. Since the imaginary part is clearly always negative, the modes are stable. In addition, we have computed the quasinormal modes using the 6th order WKB method, and we find an excellent agreement between the semi-analytical results and our exact analytical expression.
Finally, the impact of the cosmological constant and of the non-minimal coupling on the spectrum is briefly discussed.


\begin{acknowledgments}

G. P. is grateful to the anonymous reviewers for comments and suggestions that improved the quality of the manuscript. The author thanks the Funda\c c\~ao para a Ci\^encia e Tecnologia (FCT), Portugal, for the financial support to the Center for Astrophysics and Gravitation-CENTRA,  Instituto Superior T\'ecnico,  Universidade de Lisboa,  through the Grant No. UID/FIS/00099/2013.

\end{acknowledgments}


\end{document}